\documentclass[a4paper]{spie}  %>>> use this instead for A4 paper
\def\aj{\ref@jnl{AJ}}                   % Astronomical Journal
\def\actaa{\ref@jnl{Acta Astron.}}      % Acta Astronomica
\def\araa{\ref@jnl{ARA\&A}}             % Annual Review of Astron and Astrophys
\def\apj{\ref@jnl{ApJ}}                 % Astrophysical Journal
\def\apjl{\ref@jnl{ApJ}}                % Astrophysical Journal, Letters
\def\apjs{\ref@jnl{ApJS}}               % Astrophysical Journal, Supplement
\def\ao{\ref@jnl{Appl.~Opt.}}           % Applied Optics
\def\apss{\ref@jnl{Ap\&SS}}             % Astrophysics and Space Science
\def\aap{\ref@jnl{A\&A}}                % Astronomy and Astrophysics
\def\aapr{\ref@jnl{A\&A~Rev.}}          % Astronomy and Astrophysics Reviews
\def\aaps{\ref@jnl{A\&AS}}              % Astronomy and Astrophysics, Supplement
\def\azh{\ref@jnl{AZh}}                 % Astronomicheskii Zhurnal
\def\baas{\ref@jnl{BAAS}}               % Bulletin of the AAS
\def\bac{\ref@jnl{Bull. astr. Inst. Czechosl.}}
\def\caa{\ref@jnl{Chinese Astron. Astrophys.}}
\def\cjaa{\ref@jnl{Chinese J. Astron. Astrophys.}}
\def\icarus{\ref@jnl{Icarus}}           % Icarus
\def\jcap{\ref@jnl{J. Cosmology Astropart. Phys.}}
\def\jrasc{\ref@jnl{JRASC}}             % Journal of the RAS of Canada
\def\memras{\ref@jnl{MmRAS}}            % Memoirs of the RAS
\def\mnras{\ref@jnl{MNRAS}}             % Monthly Notices of the RAS
\def\na{\ref@jnl{New A}}                % New Astronomy
\def\nar{\ref@jnl{New A Rev.}}          % New Astronomy Review
\def\pra{\ref@jnl{Phys.~Rev.~A}}        % Physical Review A: General Physics
\def\prb{\ref@jnl{Phys.~Rev.~B}}        % Physical Review B: Solid State
\def\prc{\ref@jnl{Phys.~Rev.~C}}        % Physical Review C
\def\prd{\ref@jnl{Phys.~Rev.~D}}        % Physical Review D
\def\pre{\ref@jnl{Phys.~Rev.~E}}        % Physical Review E
\def\prl{\ref@jnl{Phys.~Rev.~Lett.}}    % Physical Review Letters
\def\pasa{\ref@jnl{PASA}}               % Publications of the Astron. Soc. of Australia
\def\pasp{\ref@jnl{PASP}}               % Publications of the ASP
\def\pasj{\ref@jnl{PASJ}}               % Publications of the ASJ
\def\rmxaa{\ref@jnl{Rev. Mexicana Astron. Astrofis.}}%
\def\qjras{\ref@jnl{QJRAS}}             % Quarterly Journal of the RAS
\def\skytel{\ref@jnl{S\&T}}             % Sky and Telescope
\def\solphys{\ref@jnl{Sol.~Phys.}}      % Solar Physics
\def\sovast{\ref@jnl{Soviet~Ast.}}      % Soviet Astronomy
\def\ssr{\ref@jnl{Space~Sci.~Rev.}}     % Space Science Reviews
\def\zap{\ref@jnl{ZAp}}                 % Zeitschrift fuer Astrophysik
\def\nat{\ref@jnl{Nature}}              % Nature
\def\iaucirc{\ref@jnl{IAU~Circ.}}       % IAU Cirulars
\def\aplett{\ref@jnl{Astrophys.~Lett.}} % Astrophysics Letters
\def\apspr{\ref@jnl{Astrophys.~Space~Phys.~Res.}}
\def\bain{\ref@jnl{Bull.~Astron.~Inst.~Netherlands}} 
\def\fcp{\ref@jnl{Fund.~Cosmic~Phys.}}  % Fundamental Cosmic Physics
\def\gca{\ref@jnl{Geochim.~Cosmochim.~Acta}}   % Geochimica Cosmochimica Acta
\def\grl{\ref@jnl{Geophys.~Res.~Lett.}} % Geophysics Research Letters
\def\jcp{\ref@jnl{J.~Chem.~Phys.}}      % Journal of Chemical Physics
\def\jgr{\ref@jnl{J.~Geophys.~Res.}}    % Journal of Geophysics Research
\def\jqsrt{\ref@jnl{J.~Quant.~Spec.~Radiat.~Transf.}}
\def\memsai{\ref@jnl{Mem.~Soc.~Astron.~Italiana}}
\def\nphysa{\ref@jnl{Nucl.~Phys.~A}}   % Nuclear Physics A
\def\physrep{\ref@jnl{Phys.~Rep.}}   % Physics Reports
\def\physscr{\ref@jnl{Phys.~Scr}}   % Physica Scripta
\def\planss{\ref@jnl{Planet.~Space~Sci.}}   % Planetary Space Science
\def\procspie{\ref@jnl{Proc.~SPIE}}   % Proceedings of the SPIE

\usepackage[]{graphicx}
\usepackage[]{subfigure}
\usepackage[]{url}

\title{Simulations of the X-ray imaging capabilities of the Silicon Drift Detectors (SDD) for the LOFT Wide Field Monitor}

\author{Y.~Evangelista\supit{a,b}, R.~Campana\supit{a,b}, E.~Del~Monte\supit{a,b}, I.~Donnarumma\supit{a}, M.~Feroci\supit{a,b}, F.~Muleri\supit{a,b}, L.~Pacciani\supit{a,b}, P.~Soffitta\supit{a,b},
A.~Rachevski\supit{c}, A.~Vacchi\supit{c}, G.~Zampa\supit{c}, N.~Zampa\supit{c}, 
S.~Suchy\supit{d},
S.~Brandt\supit{e}, C.~Budtz-J{\o}rgensen\supit{e} and M.~Hernanz\supit{f}
\skiplinehalf
\supit{a}INAF/IAPS Rome, Via del Fosso del Cavaliere 100, I-00133, Rome, Italy;\\
\supit{b}INFN--Roma 2, Via della Ricerca Scientifica 1, I-00133, Rome, Italy;\\
\supit{c}INFN--Trieste, Padriciano 99, I-34127, Trieste, Italy;\\
\supit{d}Institute for Astronomy and Astrophysics, Eberhard Karls University T\"ubingen, Sand 1, 72076, Tuebingen, Germany; \\
\supit{e}Technical University of Denmark -- DTU SPACE, National Space Institute, Elektrovej Building 327, DK-2800, Kgs. Lyngby, Denmark;\\
\supit{f}IEEC/CSIC, Campus UAB, Fac. Ci\`{e}ncies Torre C5 parell 2, ES-08193, Bellaterra, Spain.
}

\authorinfo{E-mail: yuri.evangelista@iaps.inaf.it, Telephone: +39 06 4993 4657}
\begin{document} 
  \maketitle 

%%%%%%%%%%%%%%%%%%%%%%%%%%%%%%%%%%%%%%%%%%%%%%%%%%%%%%%%%%%%% 
\begin{abstract}

The Large Observatory For X-ray Timing (LOFT), selected by ESA as one of the four Cosmic Vision M3 candidate missions to 
undergo an assessment phase, will revolutionize the study of compact objects in our galaxy and of the brightest supermassive black holes in active galactic nuclei. 
The Large Area Detector (LAD), carrying an unprecedented effective area of 10~m$^2$, is complemented by a coded-mask Wide Field Monitor, in charge of monitoring a large 
fraction of the sky potentially accessible to the LAD, to provide the history and context for the sources observed by LAD and to trigger its observations on their most 
interesting and extreme states. In this paper we present detailed simulations of the imaging capabilities of the Silicon Drift Detectors developed for the 
LOFT Wide Field Monitor detection plane. The simulations explore a large parameter space for both the detector design and the environmental conditions, 
allowing us to optimize the detector characteristics and demonstrating the X-ray imaging performance of the large-area SDDs in the 2--50~keV energy band.

\end{abstract}

%>>>> Include a list of keywords after the abstract 


%%%%%%%%%%%%%%%%%%%%%%%%%%%%%%%%%%%%%%%%%%%%%%%%%%%%%%%%%%%%%
\section{INTRODUCTION}
\label{sec:intro}  % \label{} allows reference to this section

%High-time-resolution X-ray observations of compact objects provide direct access to strong-field gravity, black hole masses and spins, and the equation of state of ultradense matter.  A 10 m$^2$-class instrument in combination with good spectral resolution is required to exploit the relevant diagnostics and answer two fundamental questions of ESA’s Cosmic Vision Theme “Matter under extreme conditions”, namely: does matter orbiting close to the event horizon follow the predictions of general relativity? What is the equation of state of matter in neutron stars? Thanks to an innovative design and the development of large monolithic silicon drift detectors, the Large Area Detector (LAD) on board the Large Observatory For x-ray Timing (LOFT) achieves an effective area of $>$10~m$^2$ (more than an order of magnitude larger than current spaceborne X-ray detectors) in the 2--30 keV range (up to 50 keV in expanded mode), yet still fits a conventional platform and small/medium-class launcher. With this large area and a spectral resolution of  $<$260~eV (@ 6 keV), LOFT will revolutionise the study of collapsed objects in our galaxy and of the brightest supermassive black holes in active galactic nuclei, yielding unprecedented information on strongly curved spacetimes and matter under extreme conditions of density and pressure.\\

The LOFT\cite{Feroci2011} WFM is a coded aperture imaging experiment designed on the heritage of the SuperAGILE experiment 
\cite{Feroci2007}, successfully operating in orbit since 2007 \cite{Feroci2010}. 
With the $\sim$120~$\mathrm{\mu}$m position resolution provided by its single-sided Silicon microstrip detector, SuperAGILE demonstrated the 
feasibility of a compact, large-area, light, low-power and high resolution X-ray imager, with steradian-wide field of view. 
The LOFT WFM applies the same concept, with improvements provided by the superior performance 
Silicon Drift Detectors (SDDs) in place of the Si microstrips. These detectors provide a lower energy threshold and better energy resolution with respect
to the microstrips, and represent the state-of-the art in large area, monolithic Silicon detectors.

The working principle of the WFM is the classical sky encoding by coded masks\cite{Fenimore1978}
and is widely used in space borne instruments (e.g. INTEGRAL, RXTE/ASM, Swift/BAT). The mask shadow recorded by the 
position-sensitive detector can be deconvolved by using the procedures\cite{Fenimore1981} to recover the image of the sky, 
with an angular resolution that depends on the mask element pitch, the mask-detector distance and the detector spatial resolution.
In this paper we describe the results obtained by means of a Monte Carlo simulator specifically developed to evaluate the spatial resolution
capabilities of the large area Silicon Drift Detectors. This tool allowed us to understand and optimize the detector performance by investigating
different detector designs (i.e. anode pitch), environmental conditions (detector temperature, radiation damage) and read-out electronics performance.

In the following we will describe the SDD working principle and the charge cloud dynamics inside the Silicon bulk (Section~\ref{sec:working_principle}), the preliminary
imaging results obtained with a 8-channels discrete read-out (Section~\ref{sec:ALICE_heritage}), the structure of the MC simulator developed for the 
LOFT/WFM (Section~\ref{sec:simulator}) and the simulation results (Section~\ref{sec:results}). 
Finally, in Section~\ref{sec:conclusions}, we draw our conclusions.

%%%%%%%%%%%%%%%%%%%%%%%%%%%%%%%%%%%%%%%%%%%%%%%%%%%%%%%%%%%%%
%%%%%%%% FIGURA PRINCIPIO DI FUNZIONAMENTO SDD
\begin{figure}[htbp]
\centering
\subfigure[]{\includegraphics[width=0.55\textwidth]{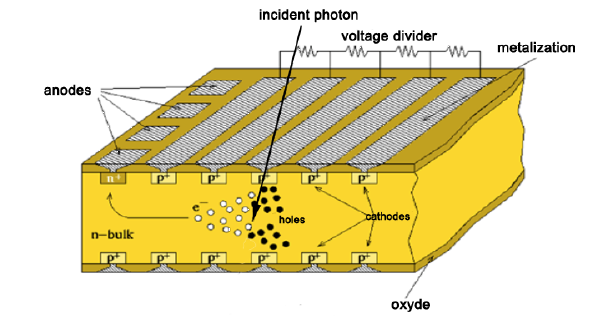}}
\subfigure[]{\includegraphics[width=0.40\textwidth]{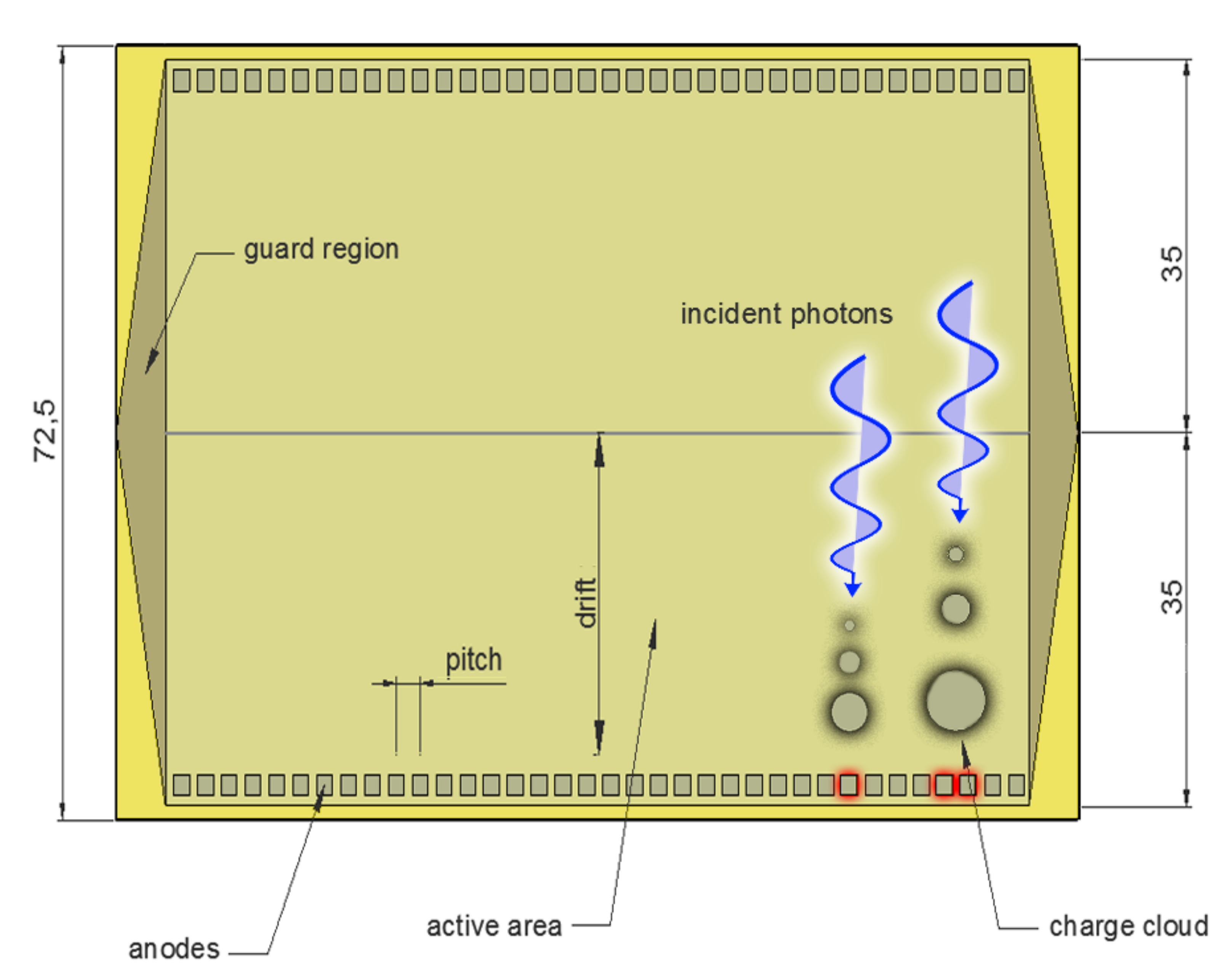}}
\caption{Working principle of the large area Silicon drift detectors}
\label{fig:sdd_wp}
\end{figure}
%%%%%%%%
\section{Working principle of the Silicon Drift Detector} 
\label{sec:working_principle}
As reported above, the proposed technology for the WFM detection plane is represented by an improved version of the large-area Silicon Drift Detectors (SDDs) developed for the Inner Tracking System 
in the ALICE experiment of the Large Hadron Collider (LHC) at CERN\cite{Vacchi1991,Rashevsky2002}, by one of the scientific institutes in the LOFT Consortium --- INFN Trieste, Italy --- 
in co-operation with Canberra Inc.
The key properties of the Si drift detectors\cite{Gatti1984} are their capability to read-out a 
large photon collecting area with a small set of low-capacitance (thus low-noise) anodes and their very small 
weight ($\sim$1~kg~m$^{-2}$). 

The working principle of the Silicon drift detectors is shown in Figure~\ref{fig:sdd_wp}. When a photon
is absorbed in the Si bulk it creates a cloud of electron-hole pairs. The holes are collected on the cathodes implanted on both the detector surface 
while the cloud of electrons is focused in the middle plane of the detector by a parabolic potential distribution. The $e^-$cloud then 
drifts towards the read-out anodes by means of a constant electric field sustained by 
negative voltages applied to the cathodes with progressively reducing amplitude down to the anodes at 0~V. 
When focused in the detector middle plane, the cloud quickly assumes a Gaussian shape\cite{Gatti1987}. 
The diffusion in Si causes the electron cloud to expand by a factor depending 
on the temperature, the electric field and drift length, as described by the relation:
\begin{equation}
\label{eq:cloud_width}
\sigma = \sqrt{2Dt + \sigma_0^2} = \sqrt{2\frac{k_BT}{q}\mu \frac{x}{\mu  E}+\sigma_0^2} = \sqrt{2\frac{k_BT}{qE}x + \sigma_0^2}
\end{equation}
where $D$ is the diffusion coefficient, $t$ is the drift time, $k_B$ is the Boltzmann's constant, $\mathrm{\mu}$ is the electron mobility, 
$T$ is the absolute temperature, $\sigma_0$ is the initial cloud dimension and $x$ is the drift length.
In Equation~\ref{eq:cloud_width} we used the Einstein relation $D = \mu  k_B T \slash q$  to derive the final result.

The charge distribution over the collecting anodes then depends on the ($x$,$y$) absorption point in the detector.
In fact the $i$-th anode collects a charge $Q_i$ which is the integral
of the Gaussian cloud between the anode boundaries.  $Q_i$ can thus be expressed as:
\begin{equation}
\label{eq:charge_dist}
Q_i = K + \frac{Q_{\mathrm{tot}}}{2} \left[ \mathrm{Erf}\left(\frac{y_i - y_0 + p/2}{\sigma \sqrt{2}}\right) - \mathrm{Erf}\left(\frac{y_i - y_0 - p/2}{\sigma \sqrt{2}}\right)\right]
\end{equation}
where $K$ is a value, changing event by event, that represents a common baseline (i.e. a common mode noise, see [\citenum{Zampa2011}]), $Q_{\mathrm{tot}}$ is the total cloud charge, 
$y_i$ is the center of the $i$-th anode, $y_0$ is the photon absorption position in the anodic direction, $p$ is the anode pitch, $\sigma$ represents 
the standard deviation of the charge distribution and $\mathrm{Erf}$ is the error function. 

The design of the large-area SDD is illustrated in Figure~\ref{fig:sdd_wp}(b). The Si tile is electrically divided in two halves, 
with two series of read-out anodes at the edges and the highest voltage along its symmetry axis (indicated as a solid line in Figure~\ref{fig:sdd_wp}b). 
The drift length is 35~mm, and the drift field is 360~V~cm$^{-1}$, which corresponds to 1300~V maximum voltage. The drift velocity is then about 5~mm~$\mathrm{\mu}$s$^{-1}$  which translates in a
maximum drift time of $\sim$7~$\mathrm{\mu}$s. 
The maximum size of the charge cloud reaching the anodes can be estimated using Equation~\ref{eq:cloud_width} with $x=35$~mm (corresponding to an event absorbed at the bottom of the drift channel).
This leads to about 1.3~mm (97\% containment radius) at room temperature. 

%\begin{figure}[th]
%\centering
%\subfigure[]{\includegraphics[width=0.385\textwidth]{ALICE_FEE.pdf}}
%\hspace{0.5cm}
%\subfigure[]{\includegraphics[width=0.50\textwidth]{ALICE_pitch_adapter.pdf}}
%\caption{(a) ALICE detector integrated with the discrete front-end electronics. (b) The eight SF-51J JFETs mounted on the pitch adapter.}
%\label{fig:ALICE}
%\end{figure}

\begin{figure}[th]
\centering
\includegraphics[width=0.45\textwidth]{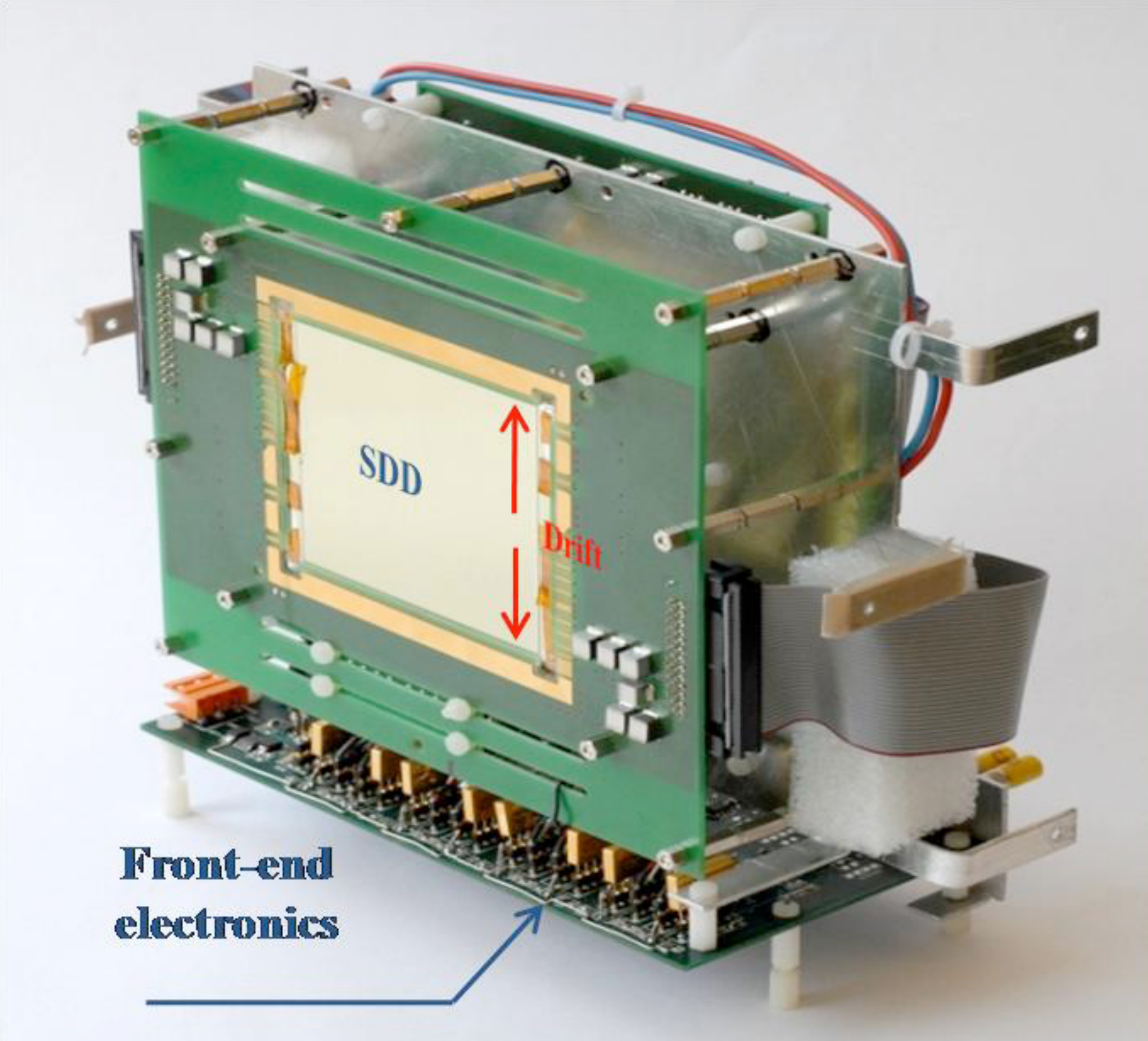}
\caption{ALICE detector integrated with the discrete front-end electronics.}
\label{fig:ALICE}
\end{figure}

\section{The ALICE heritage}
\label{sec:ALICE_heritage}
The large area multi-anode SDD was initially designed in the framework of the LHC-ALICE
experiment at CERN, and it demonstrated a spatial resolution better than 30~$\mathrm{\mu}$m in the two dimensions in localizing the impact 
point of ionizing particles\cite{Crescio2006}. 
The ALICE-D4 SDD features a sensitive area of 53~cm$^2$, with the two detector halves read-out by means of two arrays of 256 anodes with 294~$\mathrm{\mu}$m pitch.
These detectors were produced on 5-inch diameter, 300~$\mathrm{\mu}$m thick NTD (Neutron Transmutation Doped) Si wafers. 
Since 2008, 260 SDDs (for a total area of 1.37~m$^{2}$) have successfully been in operation at ALICE. 

Over the last few years R\&D work has been carried out to characterize and optimize the same detector design for detection of soft X-rays. 
The preliminary results obtained under the XDXL INFN project (in collaboration with INAF, Politecnico di Milano, Università di Pavia) 
with a spare detector of ALICE with a bread-board read-out based on discrete electronics, 
showed high spectral performance and good position resolution already at room temperature. 
A preliminary evaluation of the spectroscopic and imaging performance of the ALICE Silicon Drift Detector 
performed at the INAF/IAPS X-ray facility\cite{Muleri2008} can be found in [\citenum{Zampa2010, Zampa2011, Campana2011}].

\begin{figure}[th]
\centering
\subfigure[]{\includegraphics[width=0.45\textwidth]{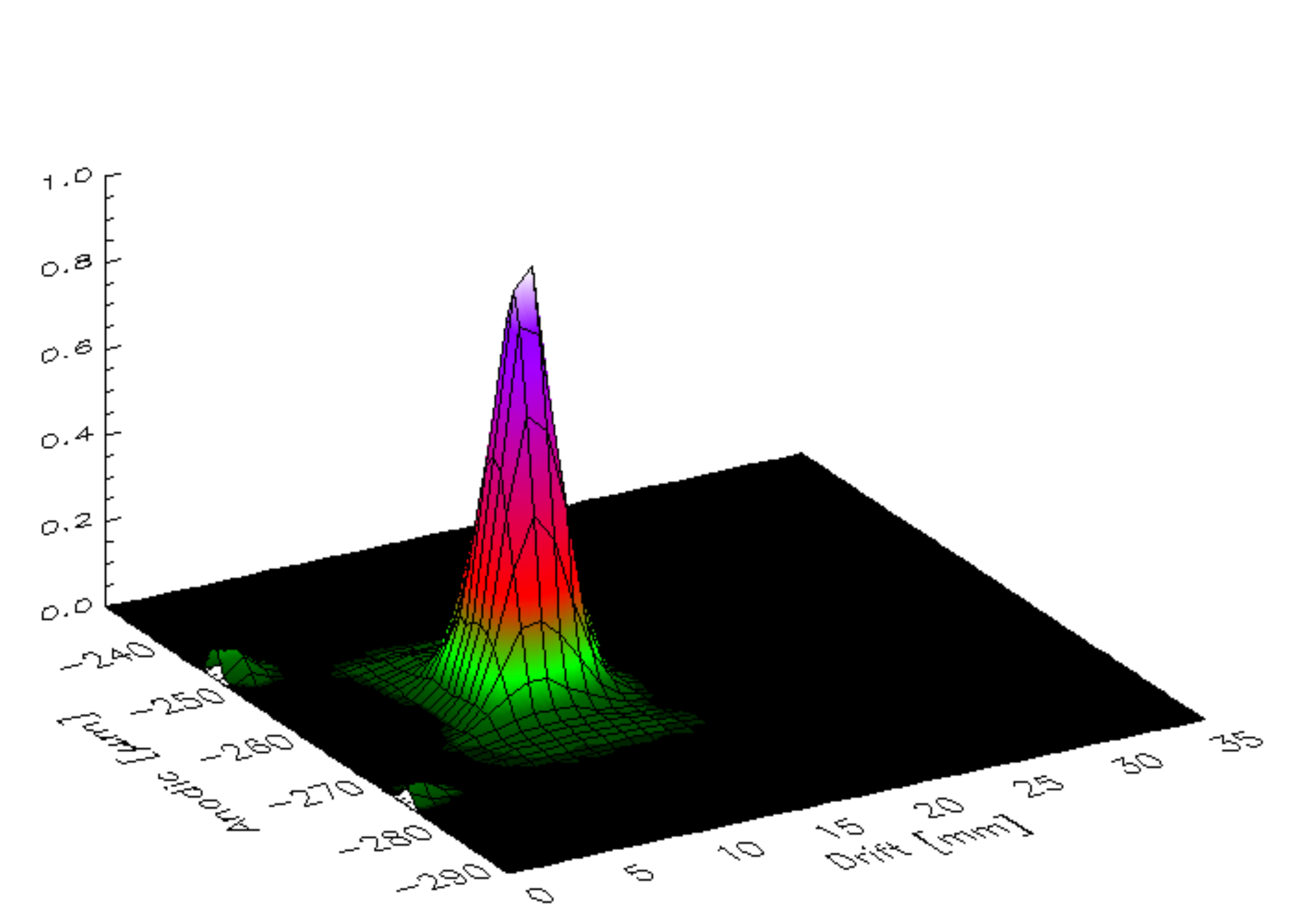}}
\subfigure[]{\includegraphics[width=0.45\textwidth]{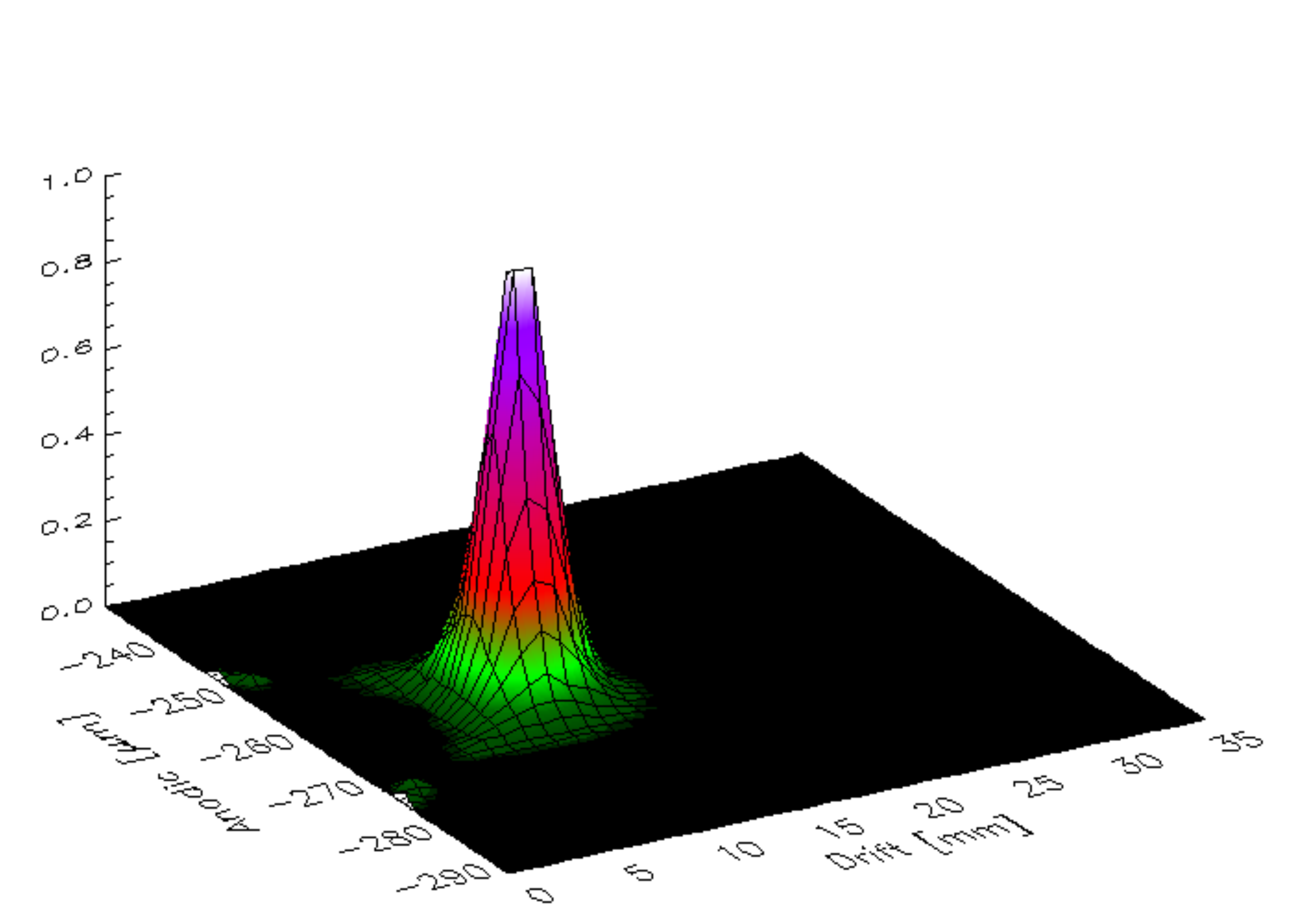}}
\caption{(a) Reconstructed detector image at T=293 K for a 1800~s integration with a 4.5~keV monochromatic X-ray beam of $\sim$100~$\mathrm{\mu}$m FWHM, at y$_0\sim0$ and x$_0\sim 10$~mm. 
(b) Simulation of a Gaussian beam with 100~$\mathrm{\mu}$m FWHM centered at y$_0=0$~mm and x$_0= 10$~mm.}
\label{fig:ALICE_imaging}
\end{figure}

Figure~\ref{fig:ALICE} shows the SDD+FEE setup used in the measurements at the INAF/IAPS laboratory. 
Eight contiguous anodes of the SDD were separately connected to the same number of low gate capacitance JFETs (C$_{GS}=0.4$~pF) used as the input 
transistor of Amptek A250F-NF charge sensitive amplifiers. 
The feedback capacitor C$_F = 50$~fF and a reset transistor are both integrated on the JFET die, allowing to reduce the input stray capacitance for a better 
noise performance. Nonetheless, some stray capacitance is unavoidable due to the need to adapt the small SDD anode pitch to the much wider one of the electronics.
%(Figure~\ref{fig:ALICE}, right panel). 
Such a parasitic capacitance was estimated to be as high as $\sim$2.5~pF with the ALICE-spare detector set-up. 
Figure~\ref{fig:ALICE_imaging}(a) shows the ALICE SDD (equipped with the discrete FEE) imaging capabilities at T$_{\mathrm{room}}$ for a 4.5~keV monochromatic X-ray beam 
(with a spot size of $\sim$100~$\mathrm{\mu}$m FWHM) as measured during the laboratory tests. 
It is worth noticing that, besides the noise performance of the discrete read-out electronics, also the small number of anodes (8) collecting the charge cloud introduces
some systematics in the determination of the Gaussian cloud parameters (Equation~\ref{eq:charge_dist}), both with respect to the spectroscopic (see Figure 7 in [\citenum{Zampa2011}])
and the imaging performance.

\section{The SDD simulator}
\label{sec:simulator}
%%%%%%%% FIGURA NUMERO DI ANODI

In order to estimate the spectroscopic and imaging capabilities of the Silicon Drift Detectors and to verify the consistence between the 
experimental results and the assumptions, we developed a Monte Carlo simulator written in IDL language and describing the charge drift and diffusion inside the SDD as well as
the detector-FEE noise characteristics.
A first version of the simulator, discussed in [\citenum{Zampa2010, Zampa2011, Campana2011}], allowed to model the ALICE-D4 SDD equipped with the 8-channels discrete FEE. 
The simulation results showed a very good agreement with the preliminary experimental measurements performed at $T_{\mathrm{room}}$, demonstrating the accuracy of the SDD modeling.
In Figure~\ref{fig:ALICE_imaging}(b) we show, as an example, the simulator output for the 8-channels discrete read-out at room temperature. This simulation was carried out
considering the same experimental conditions of the measurement showed in Figure~\ref{fig:ALICE_imaging}(a).

\begin{figure}[ht]
\centering
\subfigure[]
{\includegraphics[height=0.31\textwidth]{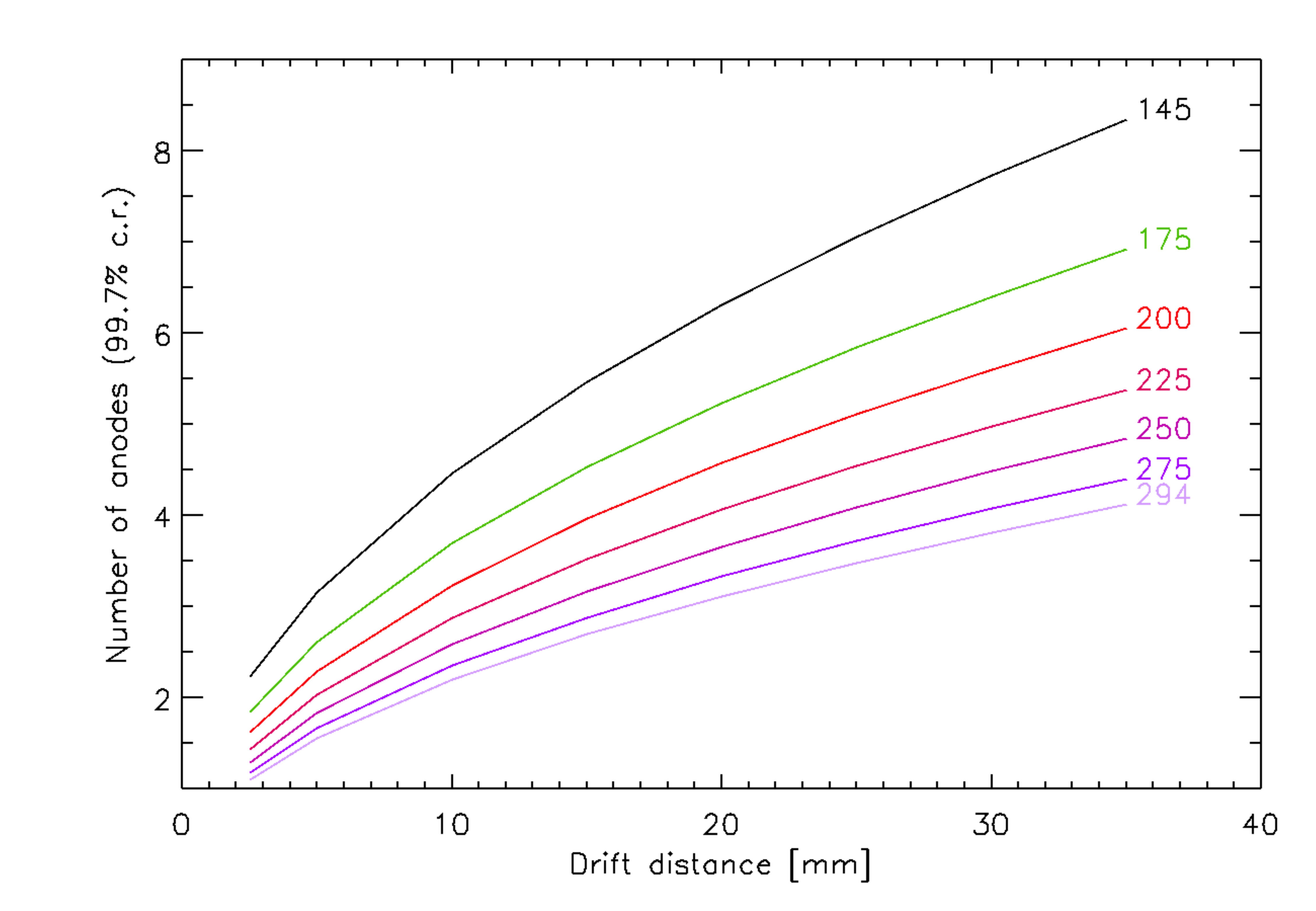}}
\hspace{0.5cm}
\subfigure[]
{\includegraphics[height=0.3\textwidth]{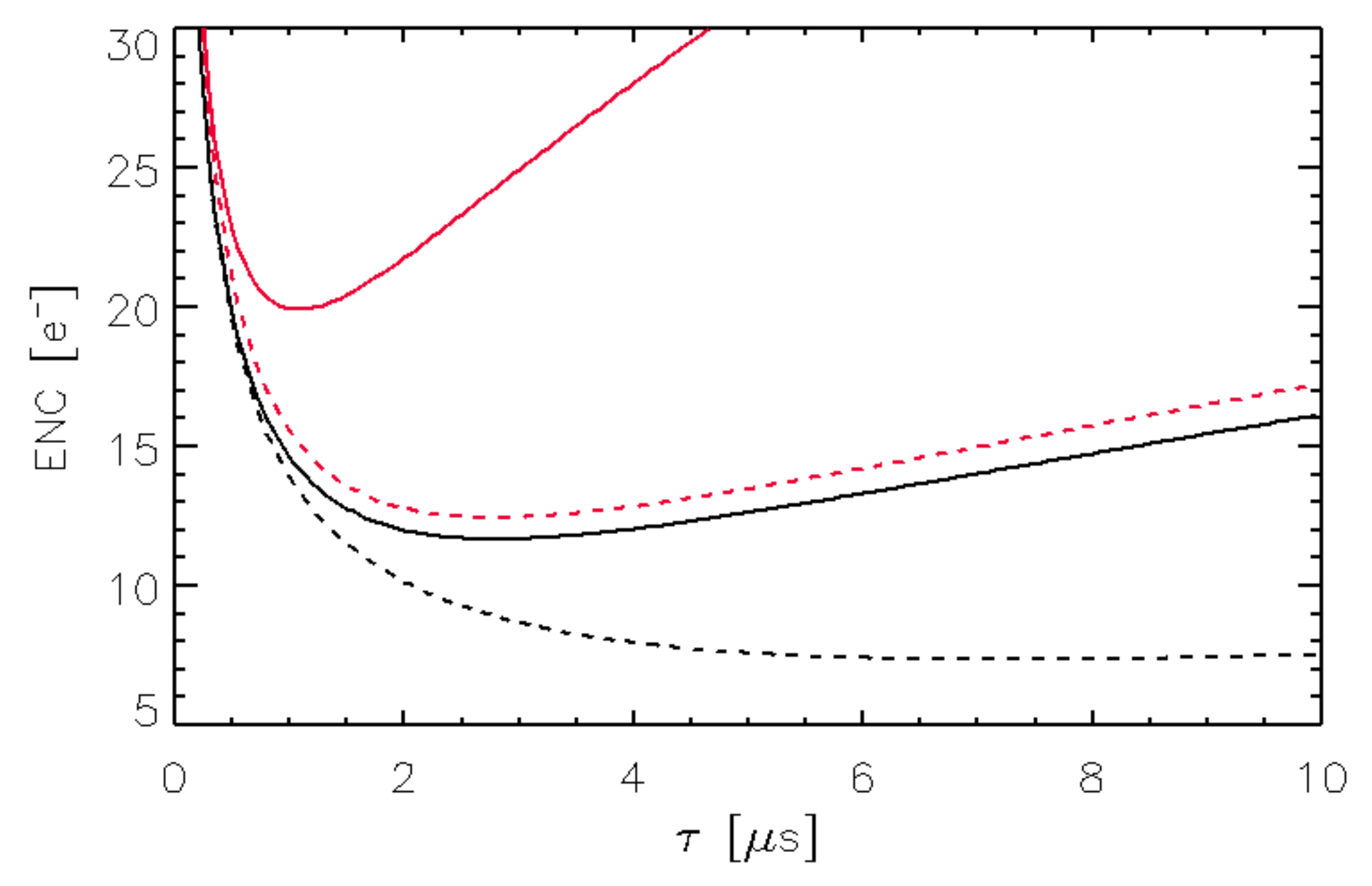}}
\caption{(a) Number of anodes collecting the electron cloud (99.7\% containment radius) as a function of the photon absorption 
point along the drift direction for anode pitches from 145~$\mathrm{\mu}$m to 294~$\mathrm{\mu}$m. 
The curves are simulated considering HV=-1300~V, T=253.15~K, $\sigma_0$=0~$\mathrm{\mu}$m. 
(b) Expected ENC at 0~$^{\circ}$C (red) and -20~$^{\circ}$C (black) for a 145~$\mathrm{\mu}$m pitch SDD in function 
of the shaping time. Curves for both Begin of Life (BOL, dashed line) and End of Life (EOL, solid line) performance are shown.}
\label{fig:anode_number_and_ENC}
\end{figure}

A new version of the simulator was then developed to study the performance obtainable by using an ASIC based FEE and by optimizing the SDD design 
(especially in regard to the anode pitch).
The estimation of the imaging (and spectroscopic) capabilities of the SDD was carried out by means of 
simulations performed at different system temperatures and anode pitches. 
For these simulations, we considered a 32--channel integrated read-out (as baselined for the LOFT/WFM ASIC) and a 
noise figure consistent with the post-layout simulations of a test ASIC being developed within the XDXL project.%LOFT collaboration.
The following scheme provides a step-by-step description of the simulator:
\begin{enumerate}
\item A photon of energy $E$ is generated in a position $x_0$ (along drift direction) and $y_0$ (along anode direction) in the SDD;
\item The photon is absorbed in the Silicon bulk and generates an electron cloud in ($x_0$,$y_0$). 
The estimation of the total charge in the cloud takes into account the average Silicon electron-hole pair generation energy (3.6~eV) and the Fano factor (0.115);
\item The cloud, which can be assumed to have an initial width of $\sigma_0\sim0$, is propagated along the drift direction, reaching the read-out anodes with a width of $\sigma$,
which is a
function of the diffusion coefficient $D$ and of the drift time $t$ as described in Equation~\ref{eq:cloud_width}. 
\item The electron cloud is then collected by a number $n$ of anodes depending on the width of the cloud and on the photon conversion point ($x_0$,$y_0$). 
The charge fraction $Q_i$ which reaches each anode is described by Equation~\ref{eq:charge_dist}. 
Figure~\ref{fig:anode_number_and_ENC}(a) shows the number of anodes collecting the electron cloud as a function of the photon absorption 
point along the drift direction for anode pitches from 145~$\mathrm{\mu}$m to 294~$\mathrm{\mu}$m with $T=253$~K and $E=360$~V~cm$^{-1}$.

\item At this stage, the simulator takes into account the Equivalent Noise Charge (ENC) for each anode independently. 
The ENC is estimated considering an integrated FEE with the following characteristics: CR--RC shaper, C$_{\mathrm{feedback}}= 21$~fF, C$_{\mathrm{stray}} =200$~fF, 
and an available power per channel of 0.722~mW.
The anode capacitance is estimated from the anode pitch and the detector thickness (between $\sim50$~fF and $\sim130$~fF for pitches in the range 145--294~$\mathrm{\mu}$m 
and 450~$\mathrm{\mu}$m thickness) while the leakage current is evaluated considering the intrinsic current measured in the laboratory 
%(I$_{\mathrm{leak}}=8640$~pA~cm$^{-3}$ at 20~$^{\circ}$C, beginning of life) 
plus the contribution due to the radiation damage after
5 years in orbit at 600~km altitude and 5$^{\circ}$ inclination, estimated by means of simulations performed with the 
ESA's Space Environment Information System (SPENVIS, \url{http://www.spenvis.oma.be}). 
The dependence of the leakage current on the temperature is taken into account by considering the relation 
\begin{equation}
I_{\mathrm{leak}}(T) \propto T^2 \cdot e^{-E_g/2 k_B T}
\label{eq:leakage}
\end{equation}
where $T$ is the absolute temperature, $q$ the electron charge, $E_g$ the Si bandgap (1.2--1.3~eV in the temperature interval $250 < T < 295$~K) and $k_B$ the Boltzmann's constant.
In Figure~\ref{fig:anode_number_and_ENC}(b) we show the expected ENC at 0~$^{\circ}$C and -20~$^{\circ}$C for a 145~$\mathrm{\mu}$m pitch as a function of the shaping time for both the 
Beginning Of Life and the End Of Life cases.
A Common Mode Noise (CMN), with an rms value of 25 e$^-$  
is then added to the 32 ASIC channels. This CMN is due to the presence of noise on the detector bias line during the laboratory tests and can be reduced by proper filtering.
\item The charge distribution is eventually fitted using Equation~\ref{eq:charge_dist}. 
The fitting parameters Q$_{\mathrm{tot}}$, y$_0$ and $\sigma$ (and $K$) allow to completely characterize the physical properties of the electron cloud (i.e. E, x$_0$, y$_0$).
\end{enumerate}

\begin{table}[!h]
\centering
\caption{Parameters used in the detector imaging simulations.}
\vspace{0.2cm}
\begin{tabular}{c|c|c|c|c}
%\hline 
\bf Anode Pitch & \bf Detector  & \bf Photon   & \bf Position along      & \bf Position along \\
\bf             & \bf Temperature & \bf Energy & \bf drift direction (x) & \bf anodic direction (y)\\
\hline 
145~$\mathrm{\mu}$m & 293.15 K & 2.0 keV & 2.5 mm    & 0 $\mathrm{\mu}$m\\
175~$\mathrm{\mu}$m & 273.15 K & 2.5 keV & 5.0 mm    & 50 $\mathrm{\mu}$m \\
200~$\mathrm{\mu}$m & 253.15 K & 3.0 keV & 10.0 mm   & 75 $\mathrm{\mu}$m  \\
225~$\mathrm{\mu}$m &          & 3.5 keV  & 15.0 mm  & 100 $\mathrm{\mu}$m \\
250~$\mathrm{\mu}$m &          & 6.0 keV  & 20.0 mm  & 150 $\mathrm{\mu}$m\\
294~$\mathrm{\mu}$m &          & 10.0 keV & 25.0 mm  &  \\
           &          & 20.0 keV  & 30.0 mm & \\
           &          &           & 32.5 mm & \\
\end{tabular}
\vspace{0.5cm}
\label{tab:parameters}
\end{table}
The simulations were performed exploring a large parameter space. Table~\ref{tab:parameters} summarizes the different values used in the simulations concerning 
the anode pitch, the detector temperature, the incident photon energy and the photon absorption point in the anodic (y) and drift (x) coordinates.

%%%%%%%%%%%%%%%%%% RISULTATI SIMULAZIONI VS PITCH

\begin{figure}[p]
\centering
\subfigure[]{\includegraphics[width=0.44\textwidth]{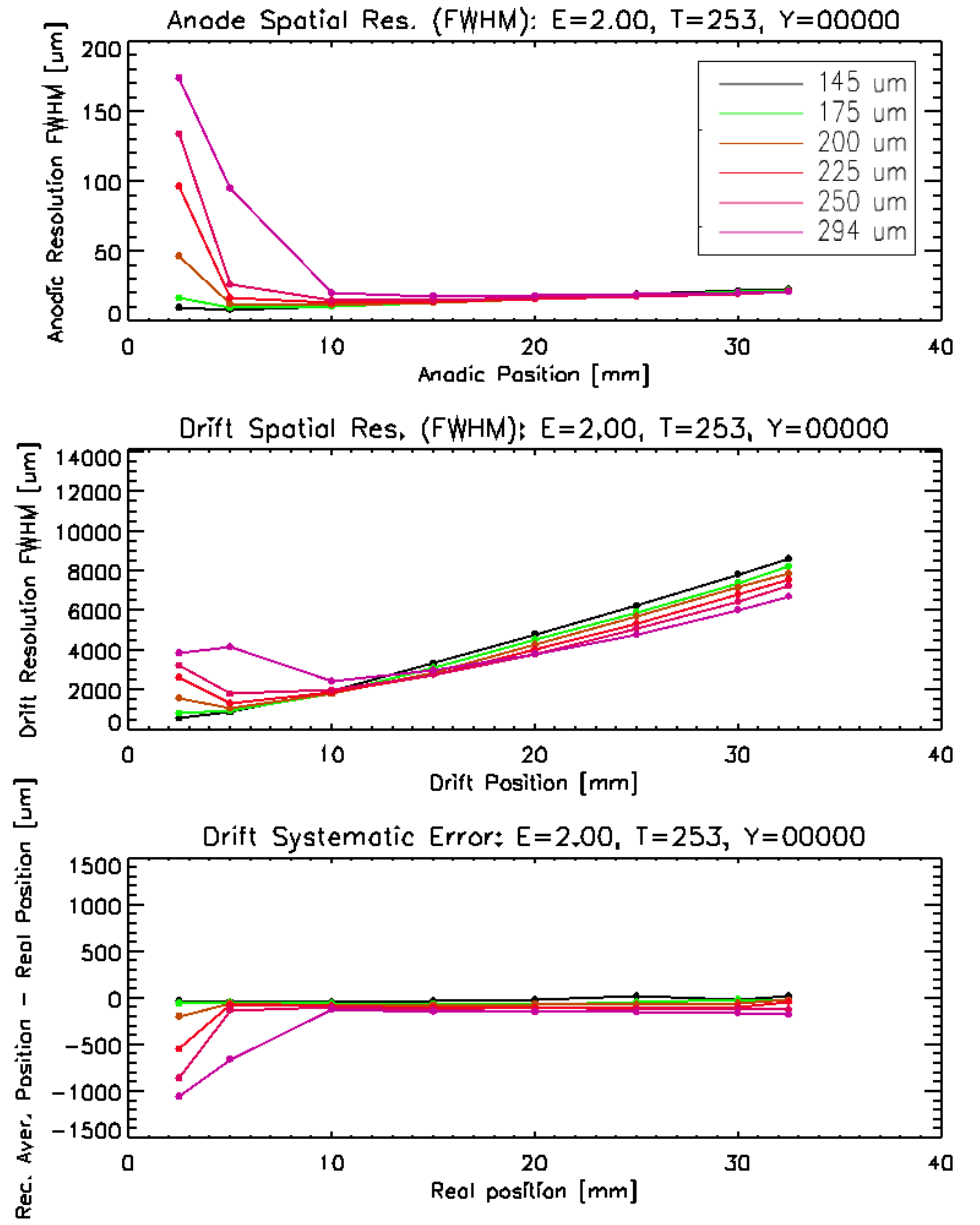}}
\subfigure[]{\includegraphics[width=0.44\textwidth]{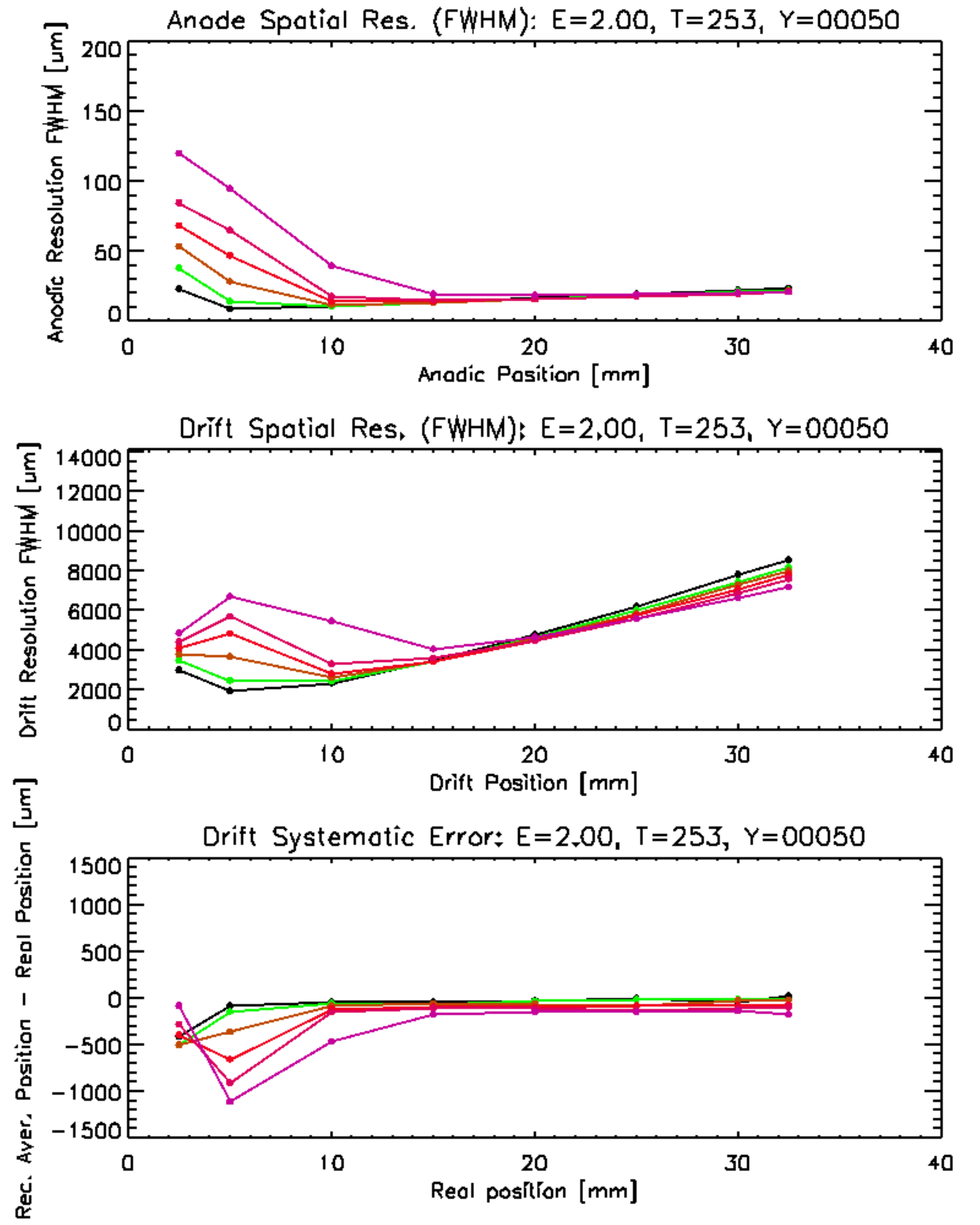}}
\vspace{0.5cm}
\subfigure[]{\includegraphics[width=0.44\textwidth]{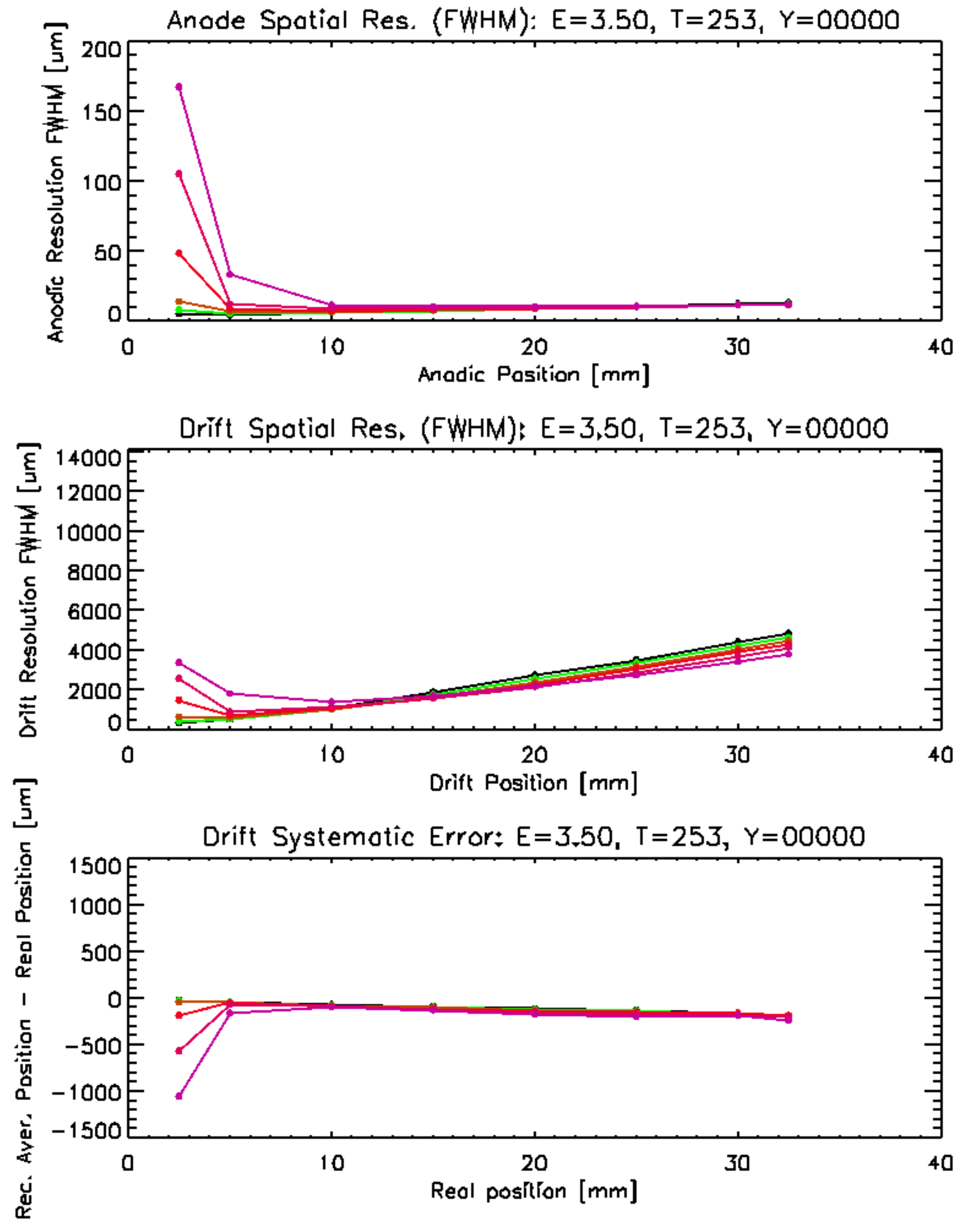}}
\subfigure[]{\includegraphics[width=0.44\textwidth]{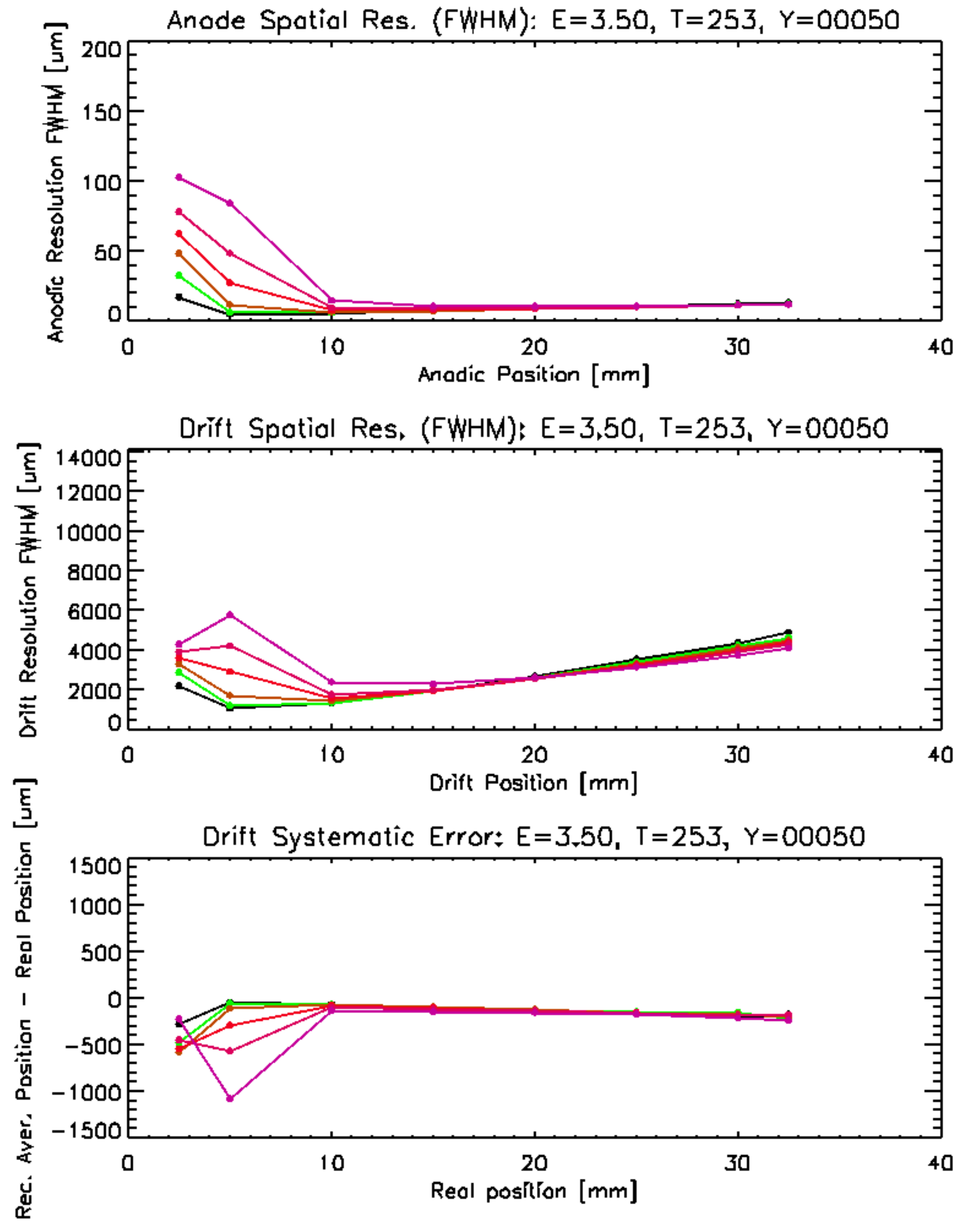}}
\caption{SDD spatial resolution ($\mathrm{\mu}$m FWHM) for E$_{\mathrm{ph}}$=2~keV (upper panels) and E$_{\mathrm{ph}}$=3.5~keV (lower panels) 
for several photon absorption positions in the anodic and drift directions and for two different positions in the anodic direction 
(left panels: y=0, right panels: y=50~$\mathrm{\mu}$m). The curves represent the resolution for
145~$\mathrm{\mu}$m (black),
175~$\mathrm{\mu}$m (green),
200~$\mathrm{\mu}$m (brown),
225~$\mathrm{\mu}$m (dark red),
250~$\mathrm{\mu}$m (light red),
and 294~$\mathrm{\mu}$m (magenta) anode pitch.
}
\label{fig:spatial_resolution_200_350}
\end{figure}

\begin{figure}[t]
\centering
\subfigure[]{\includegraphics[width=0.44\textwidth]{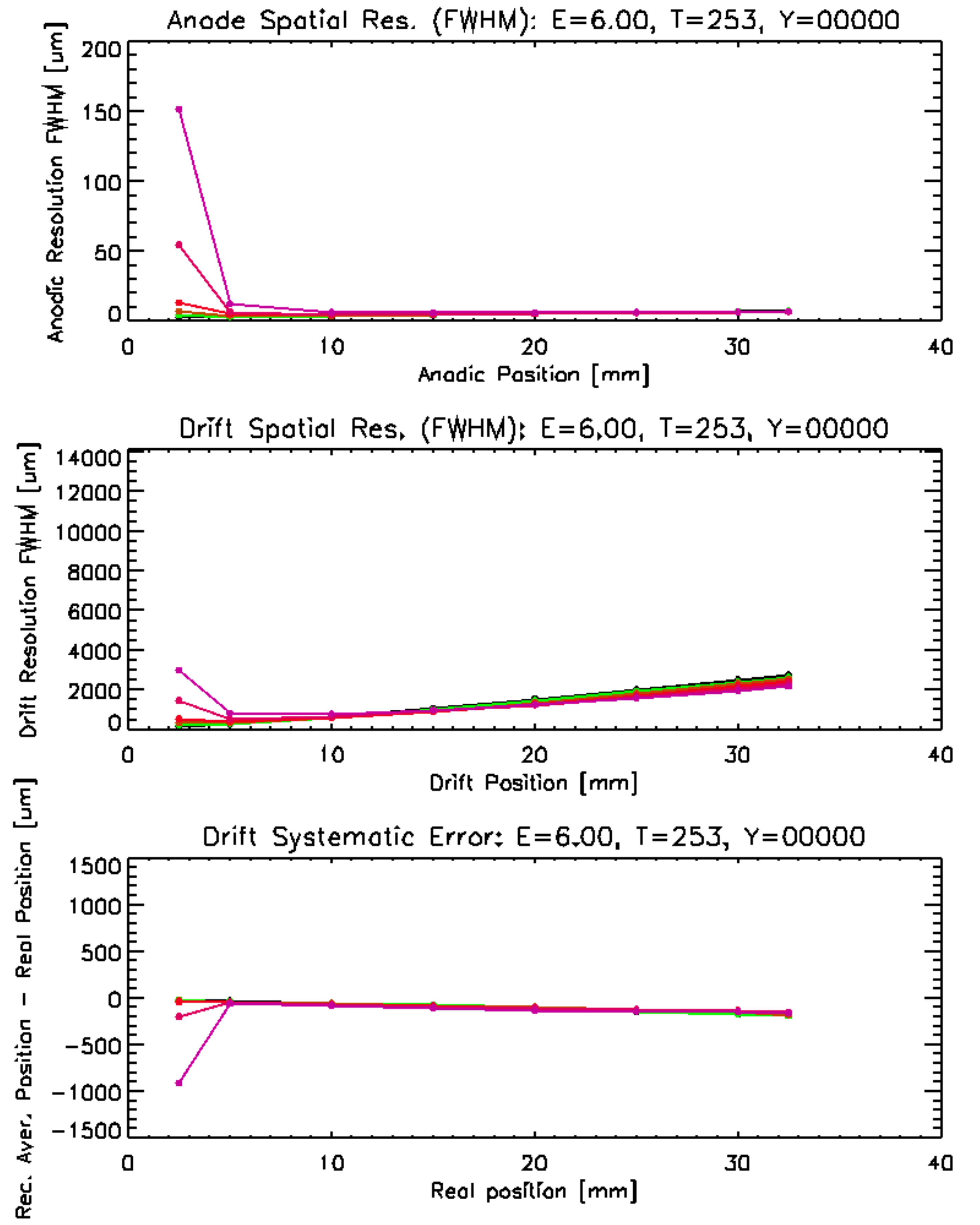}}
\subfigure[]{\includegraphics[width=0.44\textwidth]{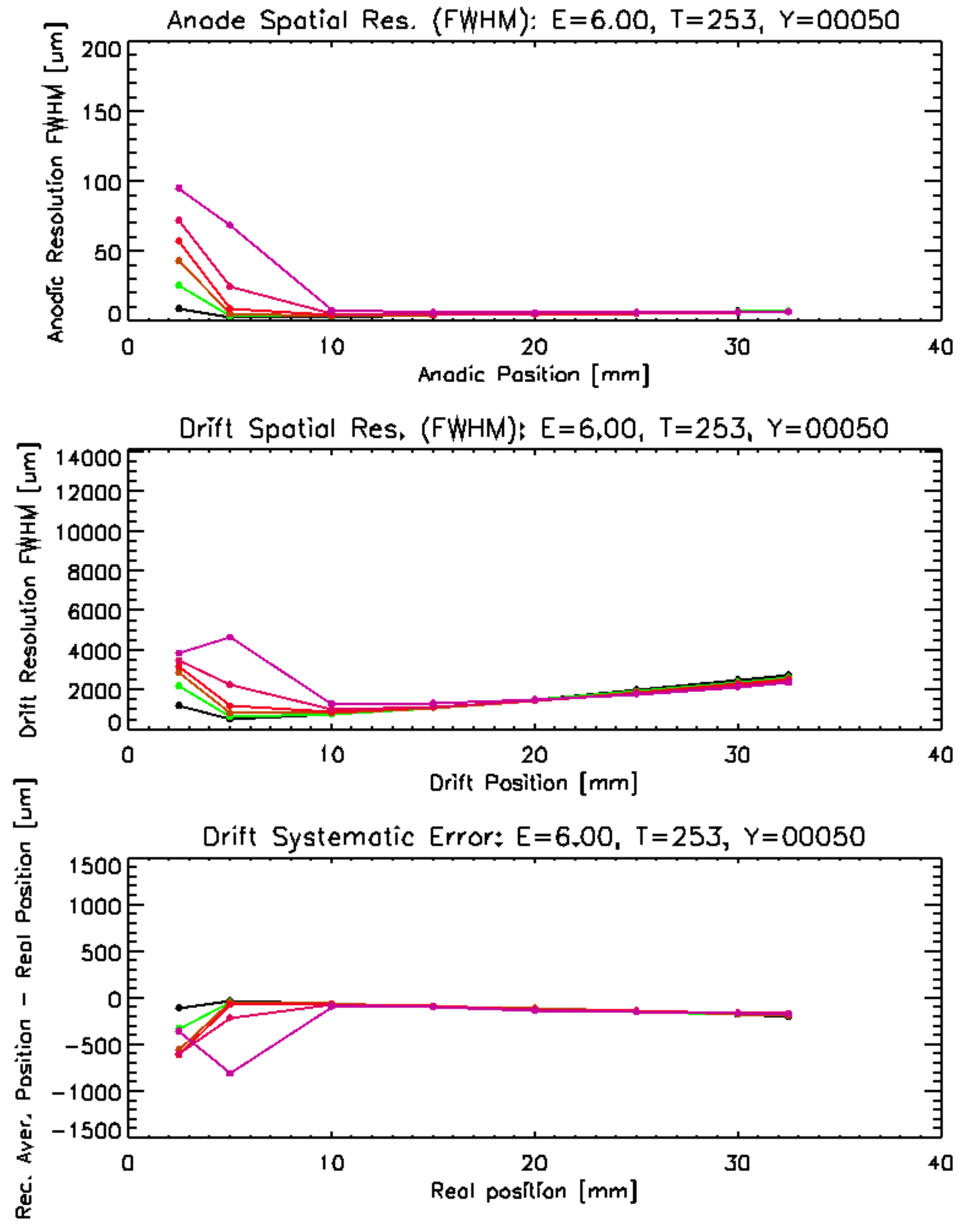}}
\caption{SDD spatial resolution ($\mathrm{\mu}$m FWHM) for E$_{\mathrm{ph}}$=6~keV for several photon absorption positions in the anodic and drift 
directions and for two different positions in the anodic direction (left panels: y=0, right panels: y=50~$\mathrm{\mu}$m). The curves represent the resolution for
145~$\mathrm{\mu}$m (black),
175~$\mathrm{\mu}$m (green),
200~$\mathrm{\mu}$m (brown),
225~$\mathrm{\mu}$m (dark red),
250~$\mathrm{\mu}$m (light red),
and 294~$\mathrm{\mu}$m (magenta) anode pitch. For energies E$\ge$6~keV, the anodic resolution becomes of the order of 
$\sim$10~$\mu$m ($x>10$~mm) for all the anode pitches considered in the simulations.}
\label{fig:spatial_resolution_600}
\end{figure}
%%%%%%%%%%%%%%%%%%%%%%%%%%%%%%%%%%%%%%%%%%%%%%%%%%%%%%%%%%%%%%%%%
\begin{figure}[t]
\centering
\subfigure[]{\includegraphics[width=0.47\textwidth]{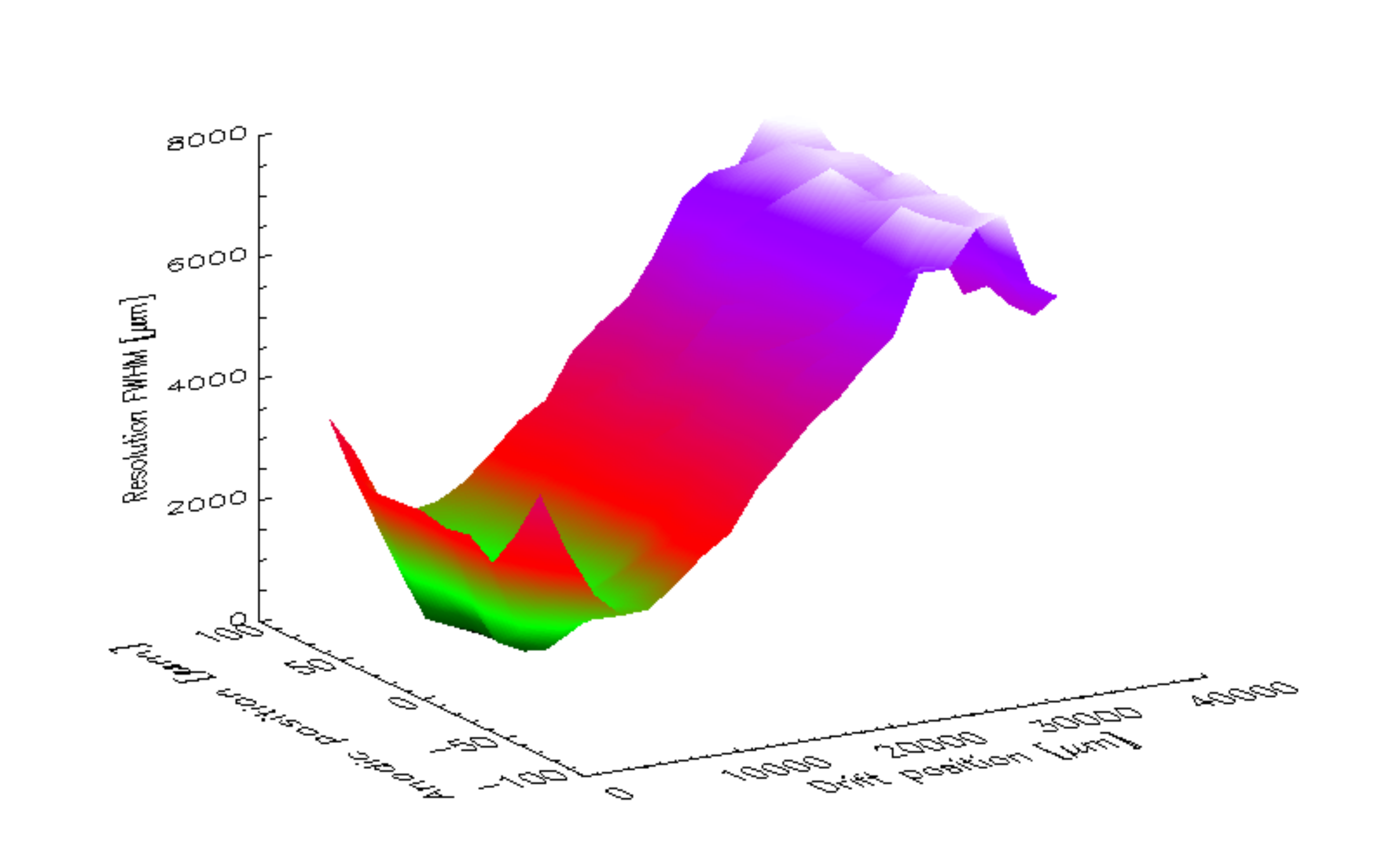}}
\subfigure[]{\includegraphics[width=0.47\textwidth]{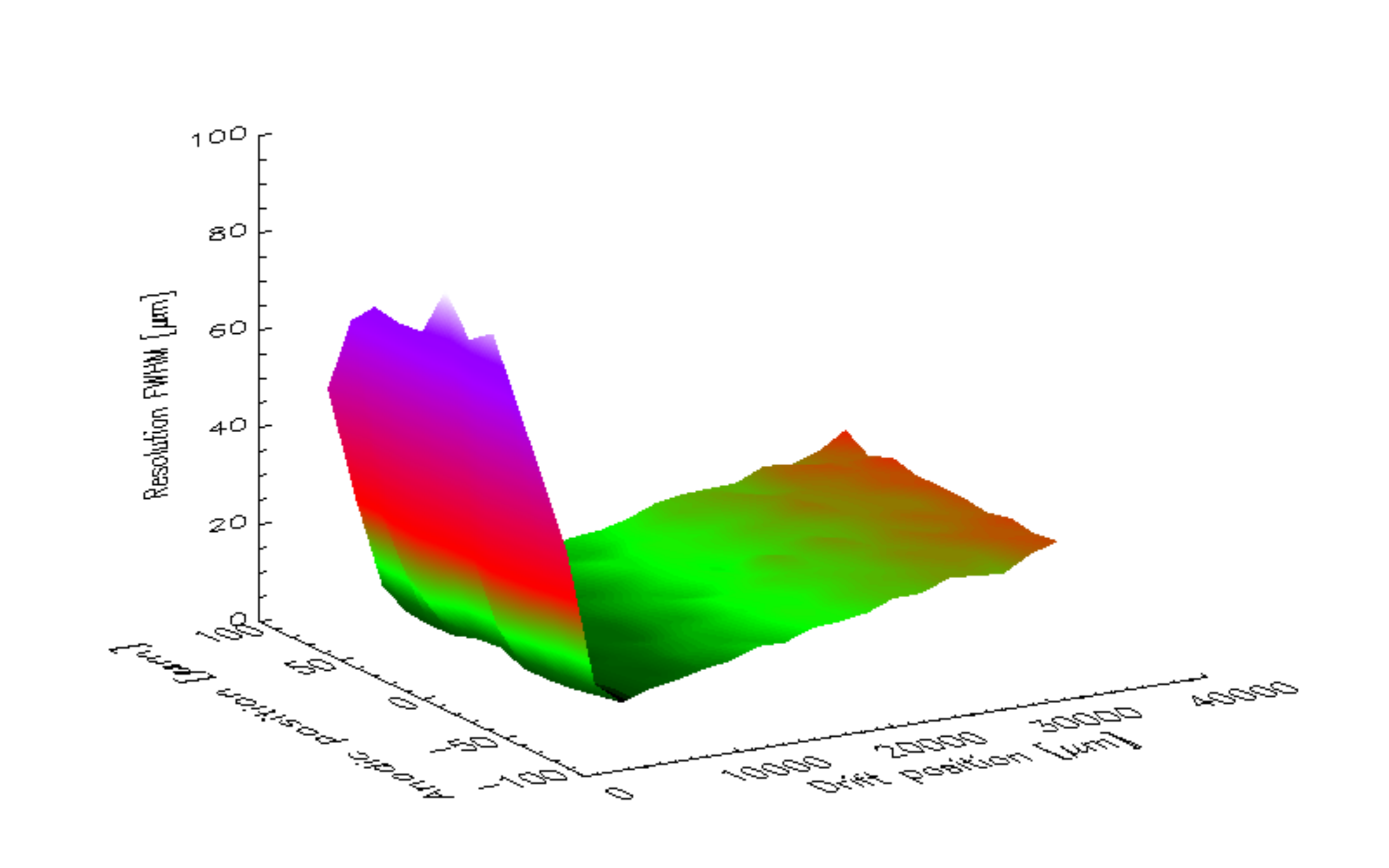}}
\caption{Drift (left) and anodic (right) spatial resolution ($\mathrm{\mu}$m FWHM) for a 2~keV photon and 145~$\mathrm{\mu}$m pitch at $T=253.15$~K. 
The maps were obtained by segmenting the area read-out by a single anode in 200 sub-pixels (14.5~$\mathrm{\mu}$m $\times$ 1.75~mm each one) and then generating 
10$^3$ events in each pixel with a uniform spatial distribution.}
\label{fig:spatial_resolution_map}
\end{figure}
%\clearpage
%%-----------------------------------------------------------

\section{Analysis of simulated data}
\label{sec:results}

For each simulation configuration we generated a large number ($2\times 10^5$) of photons. The output is saved as FITS tables,
including a truncation of the values to simulate the ADC response.
The event list containing the simulation results were then analysed fitting Equation~\ref{eq:charge_dist} by means of a least squares minimization algorithm
written in C (CMPFIT, \url{http://www.physics.wisc.edu/~craigm/idl/cmpfit.html}).

Figure~\ref{fig:spatial_resolution_200_350}(a) and Figure~\ref{fig:spatial_resolution_200_350}(b) show the anodic and 
drift spatial resolution (FWHM) for a 2~keV photon at 253~K for different anode pitches, for several photon absorption 
points in the drift direction and for two different positions in the anodic direction: at the center of the anode (y=0), and 50~$\mathrm{\mu}$m 
far from the anode center. The results shown in the plots indicate that 
a pitch smaller than 175~$\mathrm{\mu}$m is required to minimize the systematic effects on the position reconstruction 
introduced by the discretization of the charge cloud read-out and thus to optimize the detector spatial resolution at low 
energies ($<$2.5--3.0~keV). 
For energies larger than 3~keV, the Signal-to-Noise ratio in each bin of the integrated Gaussian cloud increases,
thus improving the spatial resolution as shown in Figure~\ref{fig:spatial_resolution_200_350}(c) and Figure~\ref{fig:spatial_resolution_200_350}(d) 
for E$_{\mathrm{ph}}$=3.5~keV and in Figure~\ref{fig:spatial_resolution_600} for E$_{\mathrm{ph}}$=6.0~keV.

Looking at the simulation results in Figure~\ref{fig:spatial_resolution_200_350} and Figure~\ref{fig:spatial_resolution_600}, is it clear that a
small anode pitch translates in a better spatial resolution for photons absorbed in the first part ($\sim$10--15~mm) of the drift channel. On the
other hand, the dependence of the anodic and drift resolution on the pitch is less important at the end of the drift length and almost negligible for
$15<x<25$~mm. Considering these results and the technological possibilities to build a SDD with a pitch smaller than 175~$\mathrm{\mu}$m, 
we suggest 145~$\mathrm{\mu}$m as the optimal pitch for the LOFT Wide Field Monitor.

In order to characterize the detector spatial resolution with high precision we performed a new set of simulations with a fixed anode pitch of 145~$\mathrm{\mu}$m.
In contrast to the simulations described above, this new set was performed by dividing each anode in 200 small sub-pixels, 
each with a surface of 14.5~$\mathrm{\mu}$m$\times$1.75~mm, and generating 10$^3$ photons of different energies (2.0, 2.5, 3.0, 4.5, 6.0, 10.0 and 20.0~keV) with a 
uniform spatial distribution inside the sub-pixel.
In the left panel of Figure~\ref{fig:spatial_resolution_map} we show the SDD spatial resolution FWHM (averaged on each sub-pixel) along the drift 
direction for 2~keV photons as a function of the absorption point in both the anodic and drift directions. 
In the right panel of the same figure we display the anodic spatial
resolution. Such resolution is in general of the order of 20--30~$\mathrm{\mu}$m and, as expected, becomes comparable with pitch/2 when the electron 
cloud is completely collected by one single anode.

\section{Discussion and conclusions}
\label{sec:conclusions}

We developed a Monte Carlo simulator to study the effects of the charge diffusion on the spatial resolution of the SDDs.
The simulated detector-FEE model takes into account all the most important factors which affect the detector imaging capabilities, 
investigating a number of configurations involving different detector designs and environmental conditions.
The simulations have driven the design of the SDDs for the LOFT/WFM instrument, 
demonstrating that a pitch of 145~$\mathrm{\mu}$m is the optimal choice for maximizing the detector imaging performance.
Such performance will be verified in the next months by using an ASIC-based FEE and the SDD prototypes already developed within the XDXL project in collaboration 
with Fondazione Bruno Kessler (FBK), Trento, Italy.
With this design, extremely high spatial resolution (better than 70~$\mathrm{\mu}$m FWHM in the anodic direction) can be obtained
by fitting the charge cloud distribution with a discrete Gaussian function. 
The same procedure also allows to determine the photon absorption point in the drift direction with a resolution (FWHM) better 
than 8~mm at 2 keV, 5~mm at 3.5 keV and 3~mm at 6 keV.
The Silicon Drift Detectors can thus be effectively used as a soft (E$>$2~keV) X-ray imaging detector, allowing to
perform bi-dimensional photon-by-photon imaging in the energy range 2--50~keV, with low resource requirements in term of
power, weight and complexity of the read-out electronics.

\acknowledgments     %>>>> equivalent to \section*{ACKNOWLEDGMENTS}       
The Italian team is grateful for support by ASI, INAF and INFN.
%This unnumbered section is used to identify those who have aided the authors in understanding or accomplishing the work presented and to acknowledge sources of funding.  

%%%%%%%%%%%%%%%%%%%%%%%%%%%%%%%%%%%%%%%%%%%%%%%%%%%%%%%%%%%%%
%%%%% References %%%%%

\bibliographystyle{spiebib}   %>>>> makes bibtex use spiebib.bst
\bibliography{bibliography.bib}   %>>>> bibliography data in report.bib

\end{document}